\documentclass[aps,twocolumn,pra,amsmath,amssymb,superscriptaddress,floats,nobibnotes,nofootinbib]{revtex4}

\usepackage{CJK}                     
\usepackage{amsmath}	
\usepackage{gensymb}
\usepackage{hyperref}
\hypersetup{colorlinks=true}
\usepackage{upgreek}
\usepackage{graphics}
\usepackage{hyperref}
\usepackage{epsfig}
\usepackage{color}
\usepackage{bm}
\usepackage{float}
\usepackage{ulem} 
\usepackage{dsfont}                 
\usepackage{wasysym}
\usepackage{multirow}

\usepackage{blindtext}

\usepackage{accents}

\usepackage{graphicx}

\graphicspath{{./}{./plots/}}

\usepackage{url}
\usepackage{multirow}

\newcommand{\upperRomannumeral}[1]{\uppercase\expandafter{\romannumeral#1}}

\begin{document}

\begin{CJK*}{GBK}{song}

\title{Spontaneous dimerization, spin-nematic order, and deconfined quantum critical point in a spin-$1$ Kitaev chain with tunable single-ion anisotropy}

\author{Qiang Luo}
\email[]{qiangluo@nuaa.edu.cn}
\affiliation{College of Physics, Nanjing University of Aeronautics and Astronautics, Nanjing, 211106, China}
\affiliation{Key Laboratory of Aerospace Information Materials and Physics (NUAA), MIIT, Nanjing, 211106, China}
\author{Shijie Hu}
\affiliation{Beijing Computational Science Research Center, Beijing 100084, China}
\author{Jinbin Li}
\affiliation{College of Physics, Nanjing University of Aeronautics and Astronautics, Nanjing, 211106, China}
\affiliation{Key Laboratory of Aerospace Information Materials and Physics (NUAA), MIIT, Nanjing, 211106, China}
\author{Jize Zhao}
\affiliation{School of Physical Science and Technology $\&$ Key Laboratory for Magnetism and
Magnetic Materials of the MoE, Lanzhou University, Lanzhou 730000, China}
\affiliation{Lanzhou Center for Theoretical Physics, Lanzhou University, Lanzhou 730000, China}
\author{Hae-Young Kee}
\email[]{hykee@physics.utoronto.ca}
\affiliation{Department of Physics, University of Toronto, Toronto, Ontario M5S 1A7, Canada}
\affiliation{Canadian Institute for Advanced Research, Toronto, Ontario, M5G 1Z8, Canada}
\author{Xiaoqun Wang}
\email[]{xiaoqunwang@zju.edu.cn}
\affiliation{School of Physics, Zhejiang University, Hangzhou 310058, China}

\date{\today}

\begin{abstract}
  The Kitaev-type spin chains have been demonstrated to be fertile playgrounds
  in which exotic phases and unconventional phase transitions are ready to appear.
  In this work, we use the density-matrix renormalization group method to study the quantum phase diagram
  of a spin-1 Kitaev chain with a tunable negative single-ion anisotropy (SIA).
  When the strength of the SIA is small, the ground state is revealed to be a spin-nematic phase
  which escapes conventional magnetic order but is characterized by a finite spin-nematic correlation because of the breaking spin-rotational symmetry.
  As the SIA increases, the spin-nematic phase is taken over by either a dimerized phase or an antiferromagnetic phase through an Ising-type phase transition,
  depending on the direction of the easy axis.
  For large enough SIA, the dimerized phase and the antiferromagnetic phase undergo a ``Landau-forbidden" continuous phase transition,
  suggesting new platform of deconfined quantum critical point in spin-1 Kitaev chain.
\end{abstract}


\maketitle

\section{Introduction}
The celebrated Kitaev model on the honeycomb lattice \cite{Kitaev2006} and its multitudinous variants offer unprecedented opportunities for our understanding of
exotic states of matter arising from bond-directional exchange couplings \cite{ZhangWHB2019,WangNmdLiu2019,RalkoMerino2020,LuoNPJ2021,LeeKCetal2020,GohlkeCKK2020,Luo2022PRB}
and unconventional quantum phase transitions (QPTs) that are beyond the Landau-Ginzburg-Wilson (LGW) paradigm \cite{FengZX2007,ShiYYN2009,GoJungMoon2019,LiKimKee2022}.
It is rigorously demonstrated that the ground state of the Kitaev honeycomb model is a quantum spin liquid (QSL) with fractionalized excitations
consisting of itinerant majorana fermions and localized $\mathbb{Z}_2$ vortices (visons) \cite{KnolleKCM2014}.
The quantum fluctuation can be greatly enhanced by including further nearest-neighbor interactions and off-diagonal exchanges,
giving rise to emergent phases such as the vison crystal \cite{ZhangWHB2019}, QSLs of different nature \cite{WangNmdLiu2019,RalkoMerino2020,LuoNPJ2021},
nematic paramagnet that breaks lattice rotational symmetry \cite{LeeKCetal2020,GohlkeCKK2020},
and spin-flop phase which can be interpreted as superfluid phase \cite{Luo2022PRB}.
At the same time, smoking-gun signals of the topological QPTs are observed by the change of Chern number
and the onset of the peak in the thermal Hall conductivity \cite{GoJungMoon2019,LiKimKee2022}.

While substantial efforts have been devoted to studying extended Kitaev models in two dimensions,
many intriguing phenomena regarding the collective behaviors of the excitations remain elusive
because of the numerical challenges and limitations of different computational methods.
One of the prominent examples is the antiferromagnetic (AFM) Kitaev model subject to a [111] magnetic field,
which is shown to have an intermediate region between the low-field non-Abelian QSL and the high-field polarized phase \cite{ZhuKSF2018,GohlkeMP2018,HickeyTrebst2019,PatelaTrv2019}.
The plausible perspective which asserts that the intermediate region is a gapless QSL with spinon Fermi surface has been challenged by a recent study,
where a different scenario of gapped QSL with a Chern number of 4 is proposed \cite{ZhangHalBat2022}.
Also, it is revealed by another work that the intermediate region is composed of two gapped phases with finite Chern number \cite{JiangLCQLWang2020}.
To reconcile these seemingly conflicting results, attempts have been made
on the spin-ladder analogue in which a staggered chiral phase as well as a few possible incommensurate phases appears \cite{SorensenCGK2021}
and on the spin-chain limit where a chiral soliton phase is observed \cite{SorensenGRWK2022}.
Therefore, the (quais-) one-dimensional Kitaev-type spin chains serve as fruitful grounds to offer insights into the enigmatic phases in higher dimensions.

Over the years, the Kitaev-type spin chains have been the focus of intensive research efforts since they can harbor interesting phases and unconventional QPTs \cite{SenShankar2010,AgrBrkNis2018,YangKG2020,YangJKG2020,YangSN2021,YouSunRen2020,LuoPRB2021,LuoPRR2021}.
In the Kitaev-$\Gamma$ chain where $\Gamma$ interaction is an off-diagonal exchange coupling \cite{YangKG2020},
a magnetically ordered state that displays a spin-nematic correlation occurs in the neighbor of the dominant AFM Kitaev interaction \cite{YangSN2021,LuoPRB2021}.
Thus, these studies provide a promising way towards pursuing the spin-nematic order in the models with bond-directional exchanges.
The spin-nematic state is characterized by a quadrupolar order
in which the spin-rotational symmetry is broken whereas both translational and time-reversal symmetries are retained,
constituting the magnetic analogue of liquid crystal \cite{Andreev1984,Chandra1991,Chubukov1991}.
Despite an active search for several decades, theoretical proposals of the spin-nematic order is rare
and experimental detection has been hindered by the fact that the spin-nematic order parameter is not coupled to the external magnetic field directly \cite{ManmanaPRB2011,OrlovaPRL2017}.
On the other hand, a continuous QPT between two magnetically ordered states with different symmetry breaking is reported in the Kitaev spin chain with multiple-spin interaction \cite{Macedo2022}.
Such an exotic transition is forbidden by the conventional LGW paradigm, providing another concrete example of the deconfined quantum critical point (DQCP) in one dimension \cite{Senthil2004Science}.

In contrast to the spin-$1/2$ Kitaev-type chains that have gained much attention,
the rich physics of their spin-1 counterparts remains hitherto largely unexplored.
For example, although it is revealed that the spin-1 Kitaev chain can host unusual excitations
and display an alluring double-peak structure in its specific heat \cite{LuoPRR2021},
nature of its ground state has not been understood thoroughly.
To this end, in this paper we consider a spin-1 Kitaev chain with a negative single-ion anisotropy (SIA)
whose easy axis varies from [001] direction to [110] direction, passing through the [111] direction.
We propose that the Kitaev phase is a sort of spin-nematic phase that can further be classified into two kinds,
depending on the structures of their low-lying excited states.
In the presence of an overwhelmingly dominant SIA,
we find a continuous QPT between the dimerized phase and the AFM phase which break different discrete symmetries,
showing that a DQCP is likely realized in the spin-1 Kitaev-type chain.

The remainder of the paper is constructed as follows.
In Sec.~\ref{SEC:Model} we construct the theoretical model, introduce the numerical methods, and show the resultant quantum phase diagram.
Section~\ref{SEC:IsoKG} is devoted to presenting the nature of spin-nematic phase and relevant QPTs,
which include QPTs from the dimerized (AFM) phase to the spin-nematic phase for the [001]-type ([111]-type) SIA,
behavior of the four-spin correlation function in the spin-nematic phase,
and emergence of the DQCP in the continuous dimer-AFM transition.
Finally, a brief conclusion is stated in Sec.~\ref{SEC:CONC}.

\begin{figure}[!ht]
\centering
\includegraphics[width=0.95\columnwidth, clip]{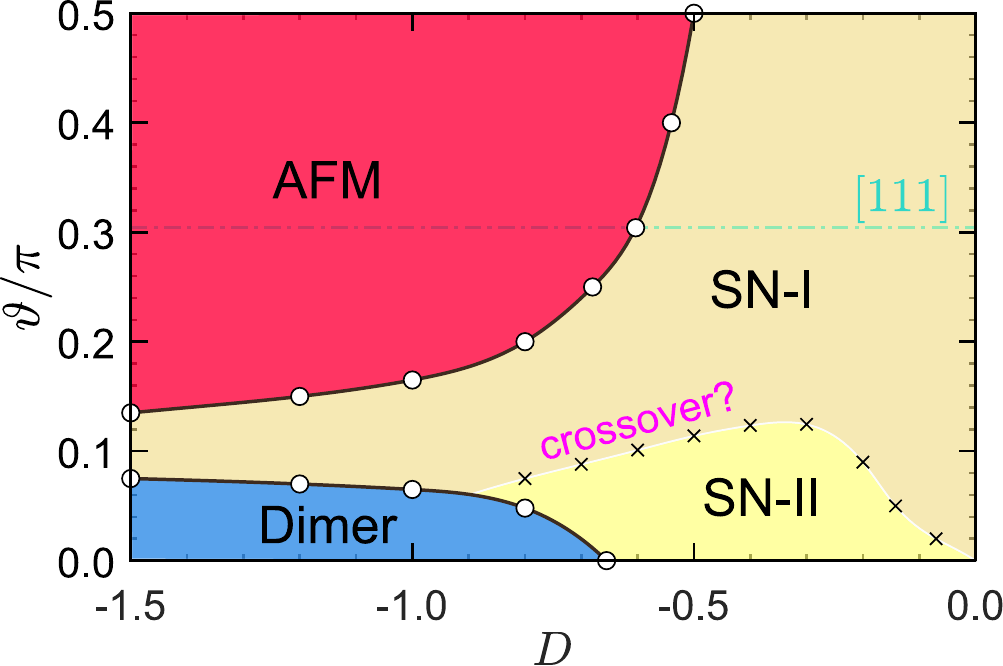}\\
\caption{Quantum phase diagram of the spin-1 Kitaev chain with a tunable SIA in which $-1.5 \leq D \leq 0$ and $0 \leq \vartheta \leq \pi/2$.
    The spin-nematic (SN) phase is gapped and can be classified into two types termed SN-I and SN-II, based on the degeneracy of their first excited states.
    They undergo a crossover rather than a phase transition as the lowest excitation gap never closes.
    The transitions from the dimerized phase and AFM phase to the spin-nematic phase belongs to the Ising universality class.
    In particular, the quantum critical point is $-0.6551(2)$ for the [001]-type SIA (i.e., $\vartheta = 0$),
    while it is $-0.6035(2)$ for the [111]-type SIA (i.e., $\vartheta = \tan^{-1}(\sqrt2)$).
    }\label{FIG-GSPD}
\end{figure}

\section{Model and Method}\label{SEC:Model}

We consider the spin-1 Kitaev chain with a tunable SIA whose Hamiltonian reads
\begin{align}\label{KtvSIA-Ham}
\mathcal{H} =& K \sum_{i=1}^{L/2} \big(S_{2i-1}^{x}S_{2i}^{x} + S_{2i}^{y}S_{2i+1}^{y}\big) \nonumber\\
&+ D \sum_{i=1}^{L} \left[\frac{\sin\vartheta}{\sqrt2}(S_i^x+S_i^y)+\cos\vartheta S_i^z\right]^2,
\end{align}
where $S_i^{\gamma}$ ($\gamma = x, y, z$) are the three components of the spin operator at the $i$th site, and $L$ is the total length of the chain which is a multiple of 4.
The first term is the Kitaev ($K$) interaction with alternating $x$- and $y$-type bonds.
The second term represents the SIA, in which $D < 0$ is the strength and $\vartheta \in [0, \pi/2]$ determines the direction of the easy axis.
The SIA term is reduced to the simple form $(S_i^z)^2$ and $(S_i^x+S_i^y)^2/2$, respectively, when $\vartheta = 0$ and $\pi/2$,
while it exhibits the form $(S_i^c)^2$ with $S_i^c = (S_i^x+S_i^y+S_i^z)/\sqrt3$ when $\vartheta = \tan^{-1}(\sqrt2) \approx 0.3041\pi$.
Although the full $SU(2)$ spin-rotational symmetry is absent,
the Hamiltonian in Eq.~\eqref{KtvSIA-Ham} respects a time-reversal symmetry $\mathcal{T}$ ($S_i^{\gamma} \mapsto -S_i^{\gamma}$)
and a link-inversion symmetry $I$ ($S_i^{\gamma} \mapsto S_{L+1-i}^{\gamma}$).
In light of a proper basis rotation
$(S_i^x, S_i^y, S_i^z)^T = \hat{R}_z\hat{R}_y \cdot (\tilde{S}_i^x, \tilde{S}_i^y, \tilde{S}_i^z)^T$
with
\begin{equation}\label{NaNb}
  \hat{R}_z =
  \left[
    \begin{array}{ccc}
      \frac{1}{\sqrt2}  &  -\frac{1}{\sqrt2}    &  0 \\
      \frac{1}{\sqrt2}  &   \frac{1}{\sqrt2}    &  0 \\
      0                 &   0                   &  1 \\
    \end{array}
  \right],
  \hat{R}_y =
  \left[
    \begin{array}{ccc}
      \cos\vartheta   &   0   &  \sin\vartheta      \\
      0             &   1   &  0                \\
      -\sin\vartheta  &   0   &  \cos\vartheta      \\
    \end{array}
  \right],
\end{equation}
it is further revealed to have a $\mathbb{Z}_2^{\tilde{x}} \times \mathbb{Z}_2^{\tilde{z}}$ dihedral symmetry $D_2$
where $\mathbb{Z}_2^{\tilde{x}/\tilde{z}}$ stands for the spin inversion in $\tilde{x}/\tilde{z}$ direction.
Due to the bond-alternating nature of the Kitaev interaction, the model possesses a two-site translational symmetry $T_2$ apparently.
However, at least in the limit cases where $\vartheta = 0$ and $\pi/2$, $\mathcal{H}$ enjoys an one-site translational symmetry $T_1$.
This can be seen by exerting the following unitary transformation on the even sites:
$(S_{2i}^x, S_{2i}^y, S_{2j}^z) \mapsto (S_{2i}^y, S_{2i}^x, -S_{2i}^z)$ \cite{SenShankar2010}.
Consequently, the Kitaev term takes the form $\sum_i S_{i}^xS_{i+1}^y$ while the SIA term remains unchanged, both of which are translationally invariant.

\begin{figure}[!ht]
\centering
\includegraphics[width=0.95\columnwidth, clip]{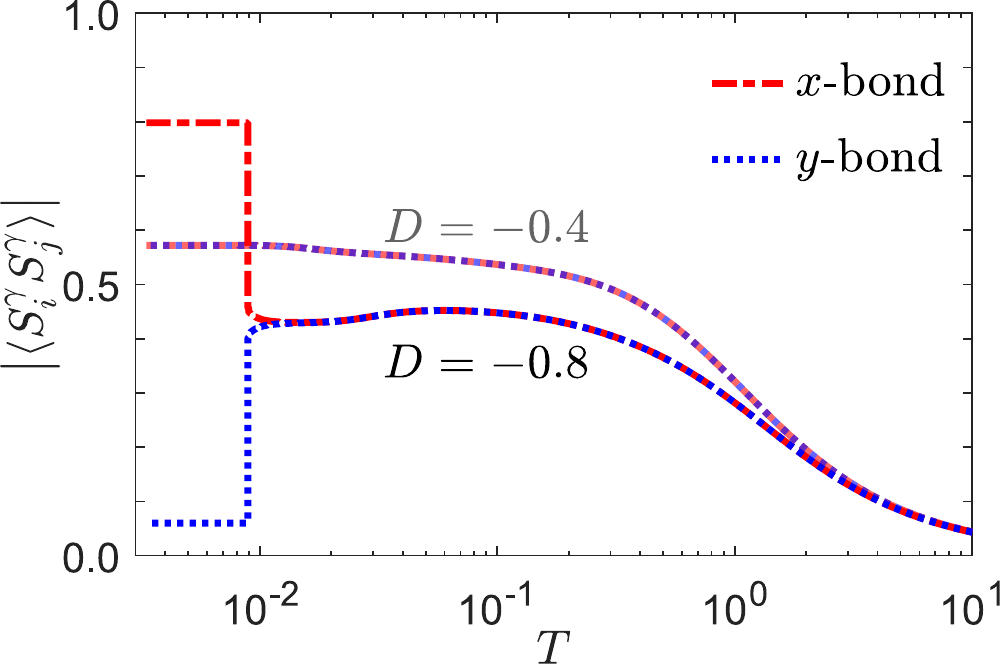}\\
  \caption{Bond strength $\big|\langle S_i^{\gamma} S_j^{\gamma}\rangle\big|$ of the $x$-bond (red dash-dot line) and $y$-bond (blue dashed line)
  as a function of temperature $T$ in the Kitaev spin chain with $\vartheta = 0.0$.
  Two different values of $D$ are chosen, which are $D = -0.4$ (in spin-nematic phase) and $D = -0.8$ (in dimerized phase).
  }\label{FIG-TMRGDimer}
\end{figure}

In fact, the SIA is naturally expected in all high-spin materials under a slight distortion from their ideal structures \cite{Stavropoulos2019PRL},
and it has been identified in various Kitaev materials like CrI$_3$, CrGeTe$_3$ and CrSiTe$_3$ \cite{Xu2018npjCM,Xu2020PRL,Stav2021PRR,Zhou2021PRB}.
Meanwhile, the role played by the [001]-type and [111]-type SIAs in the spin-1 and spin-$3/2$ Kitaev honeycomb models has been studied extensively \cite{Bradley2022PRB,Jin2022NC}.
In the large-$S$ limit, it is revealed that the SIA can stabilize an interesting triple-meron crystal consisting of three merons,
leading to a finite topological number and a quantized topological Hall conductance \cite{Chen2023NJP}.
These studies imply that Eq.~\eqref{KtvSIA-Ham} should also harbour a rich physics.

In what follows we set $K$ = 1 as the energy unit unless stated otherwise.
The quantum phase diagram is mapped out by the density-matrix renormalization group (DMRG) method \cite{White1992,Peschel1999,Schollwock2005}.
In the DMRG calculation we adopt both open (OBC) and periodic (PBC) boundary conditions alternatively,
depending on the prominent issue that matters.
To improve the numerical accuracy, 2000 block states are kept in order to maintain a small truncation error of $\sim10^{-7}$ or less.
The sweep is executed twelve times basically, with the potential to increase by several times in the vicinity of the quantum critical point.
When necessary, the transfer-matrix renormalization group (TMRG) method is also employed to study the finite-temperature evolution of physical quantities \cite{BurXiangGeh1996,WangXiang1997}.
During the calculation, we set the Trotter-Suzuki step $\tau = 0.01$ and the block states $m = 1024$.

Figure~\ref{FIG-GSPD} illustrates the quantum phase diagram in the region of $D \in [-1.5, 0.0]$ and $\vartheta \in [0, \pi/2]$ in the spin-1 Kitaev chain with tunable SIA.
Firstly, by calculating the four-spin correlation function pertaining to the spin-nematic order,
we find that the small-$D$ region, including the Kitaev limit whose ground state is previously termed Kitaev phase \cite{LuoPRR2021},
exhibits a nonzero spin-nematic correlation over the vanishing magnetic moment.
This area is thus arguably a spin-nematic phase that has long been pursued in the past decades \cite{Andreev1984,Chandra1991,Chubukov1991}.
The spin-nematic phase has a unique ground state, above which a finite excitation gap is acquired.
According to the degeneracy of its \textit{first} excited state, however, it can be further divided into two parts where a crossover occurs between them.
Secondly, the dimerized phase and the AFM phase, which break translational symmetry and dihedral and time-reversal symmetries, respectively, appear as the strength of the SIA increases.
When the strength of the SIA is moderate, the spin-nematic order is intervened between the two,
in accordance with the fact that the spin-nematic order preserves the translational symmetry and time-reversal symmetry.
Last but not the least, a continuous QPT between the dimerized phase and the AFM phase, which is advocated by a central charge of 1,
is identified if the SIA is overwhelmingly dominant.
Hence, a DQCP is likely realized in the spin-1 Kitaev-type chain.

\section{Results and Discussion}\label{SEC:IsoKG}

\subsection{Dimerized phase and AFM phase}

The dimerized phase and the AFM phase are two representative symmetry-breaking phases that have been widely recognized in the field of quantum magnetism.
For concreteness, we consider the Kitaev chain in the [001]-type ([111]-type) SIA to study the dimerized phase (AFM phase) and its transition to the spin-nematic phase.
The dimerized phase breaks the translational symmetry spontaneously, leading to a gapped ground state with a two-fold degeneracy.
In the spin-1 Heisenberg chain, the dimerized phase is demonstrated to be realized by adding competing biquadratic interaction \cite{Lauchli2006,Hu2014PRL},
three-spin interaction \cite{ChepigaAM2016}, or spatial alternation \cite{Kitazawa1996}.
Nevertheless, the SIA itself cannot induce the dimerized phase \cite{Chen2003PRB,Hu2011PRB}.

In the Kitaev chain with a [001]-type SIA, however, the intrinsic bond-directional interaction opens the possibility of realizing the dimerized phase.
Since there is only one site in each unit cell due to the translational symmetry $T_1$,
a natural way to check for the dimerized phase is by measuring the dimer order parameter defined as $O = \lim_{L\to\infty} O_L$ with
\begin{equation}\label{EQ:DimerOP}
O_{L} = \big|\langle S_{L/2-1}^x S_{L/2}^x\rangle - \langle S_{L/2}^y S_{L/2+1}^y\rangle\big|.
\end{equation}
Thus, the dimerized phase occurs as long as the bond strength of $\big|\langle S_{L/2-1}^x S_{L/2}^x\rangle\big|$ and $\big|\langle S_{L/2}^y S_{L/2+1}^y\rangle\big|$ differ.
Figure~\ref{FIG-TMRGDimer} shows the finite-temperature TMRG calculation of the bond strength
$\big|\langle S_{i}^{\gamma} S_{j}^{\gamma}\rangle\big|$ ($\gamma = x, y$) with $D = -0.4$ and $-0.8$.
As the temperature $T$ evolves from 10 to 0.0033,
the curves of the bond strength between neighboring $x$ bond (red dot-dashed line) and $y$ bond (blue dotted line) overlap persistently when $D = -0.4$.
By contrast, there is a sharp differentiation of the bond strength as long as the temperature is lower than $\sim0.01$ when $D = -0.8$,
indicating a spontaneous dimerization thereof.
In the ultra-low temperature region ($T < 0.01$), the bond strength is insensitive to the temperature
and the fact that strength of the weak bond strength remains finite down to the zero temperature reveals a partially dimerized phase.

\begin{figure}[!ht]
\centering
\includegraphics[width=0.95\columnwidth, clip]{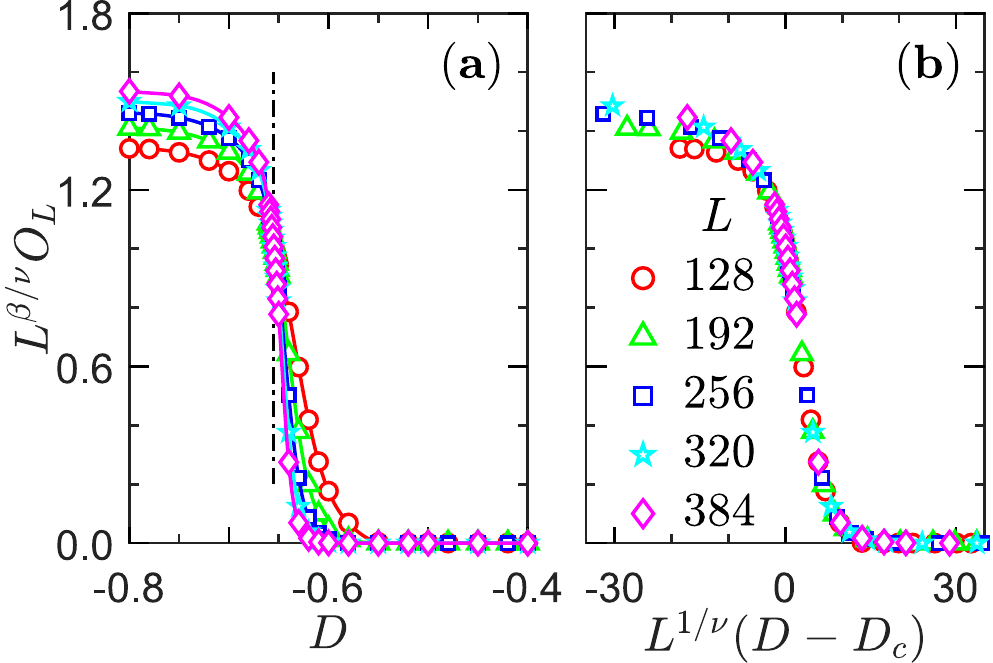}\\
  \caption{The finite-size scaling of the dimer order parameter $O_L$ as a function of $D$ in the Kitaev spin chain with $\vartheta = 0.0$.
  The chain length $L$ chosen are 128 (red circle), 192 (green triangle), 256 (blue square), 320 (cyan pentagram), and 384 (pink diamond).
  }\label{FIG-DimerFSS}
\end{figure}

To study the nature of the QPT, we use the DMRG method to calculate the dimer order parameter $O_L$ for different length $L$.
According to the finite-size scaling ansatz \cite{Fisher1972PRL}, the dimer order parameter $O_L$ satisfies the formula
\begin{equation}\label{EQ:FSS}
O_{L}(D) \simeq L^{-\beta/\nu} f_O\big(|D-D_c|L^{1/\nu}\big),
\end{equation}
where $\beta$ and $\nu$ are critical exponents of order parameter and correlation length,
and $f_O(\cdot)$ is a nonuniversal function that relies on $O_L$.
To extract the critical exponents, we adjust parameters $\mu_{1,2}$ until we see the intersection of $O_L L^{\mu_1}$ as a function of $D$
and the collapse of $O_L L^{\mu_1}$ as a function of $|D-D_c|L^{\mu_2}$ for all length $L$.
The critical exponents are then given by $\beta = \mu_1/\mu_2$ and $\nu = 1/\mu_2$.
Figure~\ref{FIG-DimerFSS} shows the finite-size scaling result of the dimer order parameter $O_L$ with $L$ = 128, 192, 256, 320, and 384.
By using of the least-square fitting method, we obtain the quantum critical point $D_c = -0.6551(2)$,
and the critical exponents $\beta = 0.123(4)$ and $\nu = 0.98(3)$.
These values are consistent with the critical exponents of the Ising transition which says that $\beta = 1/8$ and $\nu = 1$,
suggesting that the transition between the dimerized phase and the spin-nematic phase belongs to the Ising universality class.

\begin{figure}[!ht]
\centering
\includegraphics[width=0.95\columnwidth, clip]{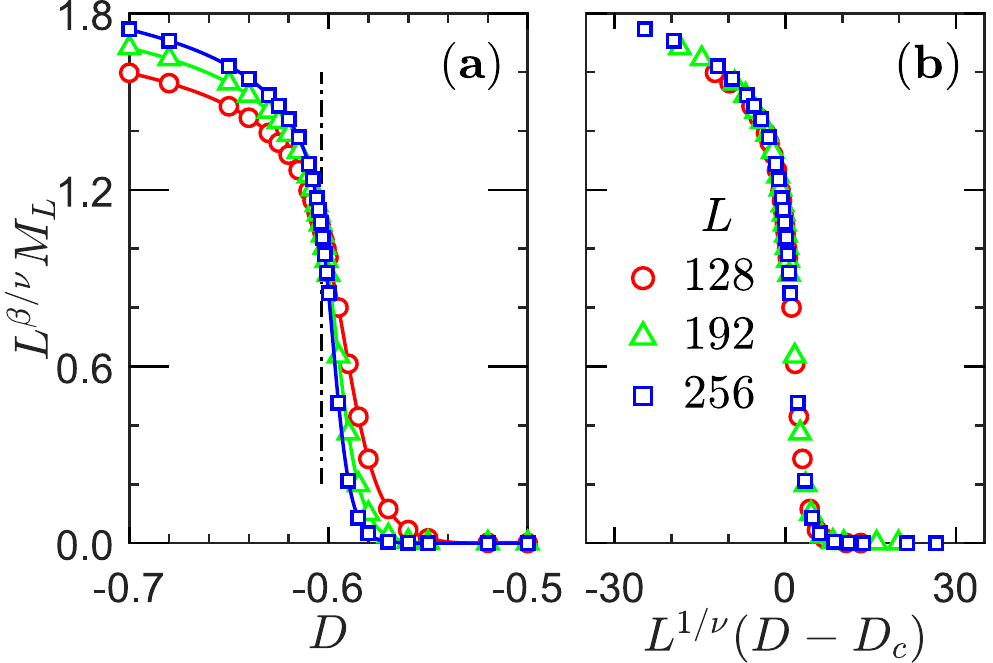}\\
  \caption{The finite-size scaling of the AFM order parameter $M_L$ as a function of $D$ in the Kitaev spin chain with $\vartheta = \tan^{-1}(\sqrt2)$.
  The chain length $L$ chosen are 128 (red circle), 192 (green triangle), and 256 (blue square).
  }\label{FIG-AFMFSS}
\end{figure}

Before proceeding further, we wish to note that the dimer order parameter in Eq.~\eqref{EQ:DimerOP} is still suitable even though the easy-axis direction of the SIA is away from the [001] direction.
After applying the local transformation $(S_{2i}^x, S_{2i}^y, S_{2j}^z) \mapsto (S_{2i}^y, S_{2i}^x, -S_{2i}^z)$,
the Kitaev interaction is translationally invariant while the SIA term takes the form
$\left[\frac{\sin\vartheta}{\sqrt2}(S_i^x+S_i^y)-(-1)^i\cos\vartheta S_i^z\right]^2$.
In the dimerized phase, $S_i^z$ is only weakly coupled to $S_i^x$ and $S_i^y$ when compared to the dominating $(S_i^z)^2$.
In addition, although the intensity of $(S_i^x)^2$ and $(S_i^y)^2$ are different, all the components of $S_i^{\alpha}S_i^{\beta}$ ($\alpha, \beta = x, y, z$) are uniformly distributed,
suggesting an effective one-site translational symmetry.

Next, we turn to study the AFM phase which is known to break the dihedral symmetry and time-reversal symmetry and exhibits a gapped doubly-degenerate ground state.
The magnetic moments along the three spin directions are all finite except for the case where $\vartheta = \pi/2$.
Due to symmetric structures of the Kitaev interaction and SIA, the $x$ and $y$ components of magnetic moments are equal but are larger than that of the $z$ component.
We apply an staggered pinning field of value $\mathcal{O}(1)$ at two end sites to slightly break the degenerate manifold.
The nondegenerate ground state thus displays a well-behaved magnetic pattern, and the magnetic order parameter can be calculated as $M = \lim_{L\to\infty} M_L$ with
\begin{equation}\label{EQ:AFMOP}
M_{L} = \sqrt{\big(\langle S_{L/2}^x\rangle\big)^2 + \big(\langle S_{L/2}^y\rangle\big)^2 + \big(\langle S_{L/2}^z\rangle\big)^2}.
\end{equation}

Figure~\ref{FIG-AFMFSS} shows the finite-size scaling result of the magnetic order parameter $M_L$ ($L$ = 128, 192, and 256) in the Kitaev chain with a [111]-type SIA.
Following a similar procedure mentioned above, we get the quantum critical point $D_c = -0.6035(2)$,
and the critical exponents $\beta = 0.127(3)$ and $\nu = 0.99(2)$,
demonstrating that the transition between the AFM phase and the spin-nematic phase also falls in the Ising universality class.

\begin{figure}[!ht]
\centering
\includegraphics[width=0.95\columnwidth, clip]{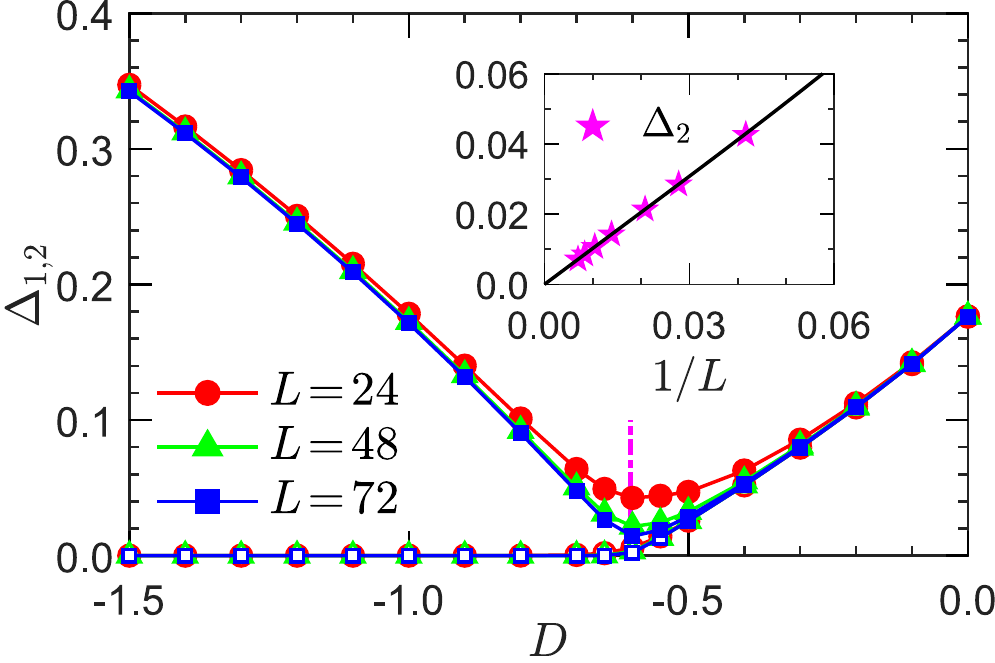}\\
  \caption{The excitation gaps $\Delta_1$ (open symbols) and $\Delta_2$ (filled symbols) as a function of $D$ in the Kitaev spin chain with $\vartheta = \tan^{-1}(\sqrt2)$.
  In the DMRG simulation, PBC are imposed and $L = 24$ (red circle), 48 (green triangle), and 72 (blue square).
  Inset: Linear extrapolation of $\Delta_2$ at the quantum critical point, with $L$ changing from 24 to 144.
  }\label{FIG-AFMGap}
\end{figure}

To further verify the continuous QPT, we calculate the lowest excitation gaps $\Delta_{1,2} = E_{1,2}-E_0$ in the vicinity of the quantum critical point.
Here, $E_{0,1,2}$ are the three lowest energy levels in the energy spectrum, with $E_0$ being the ground-state energy.
In the calculation we use the PBC to remove the boundary effect, and the ground state of the spin-nematic phase is unique while it is doubly degenerate in the AFM phase.
Behaviors of the excitation gaps $\Delta_1$ (open symbols) and $\Delta_2$ (filled symbols) as a function of $D$ are shown in Fig.~\ref{FIG-AFMGap}.
Deep in the AFM phase, $\Delta_1$ is vanishingly small and $\Delta_2$ is robust against the chain length.
As the SIA approaches the quantum critical point, the finite-size effect is significant since $\Delta_2$ decreases apparently with the increase of the system size.
The inset of Fig.~\ref{FIG-AFMGap} shows the evolution of $\Delta_2$ as a function of $1/L$ for a series of chain length $L$ ranging from 24 to 144.
The linear extrapolation gives an estimate of $0.002(5)$ for $\Delta_2$, corroborating a continuous QPT with a closure of the lowest excitation gap.

We wish to comment on the influence of the sign of the Kitaev interaction on the QPT.
For the [001]-type SIA with $\vartheta = 0$, the transformation of $(S_i^x, S_i^y, S_i^z) \mapsto (-S_i^x, -S_i^y, S_i^z)$ on all even sites
implies that $\mathcal{H}(K, D)$ = $\mathcal{H}(-K, D)$, showing that the sign of the Kitaev interaction does not alter the position of transition point.
By contrast, for the [111]-type SIA with $\vartheta = \tan^{-1}(\sqrt2)$, $\mathcal{H}(K, D)$ and $\mathcal{H}(-K, D)$ are no longer equivalent.
While the QPT is still of the Ising universality class when the Kitaev interaction is ferromagnetic,
the transition point is $-0.5531(2)$, which is larger than that of the AFM case.

\begin{figure}[!ht]
\centering
\includegraphics[width=0.95\columnwidth, clip]{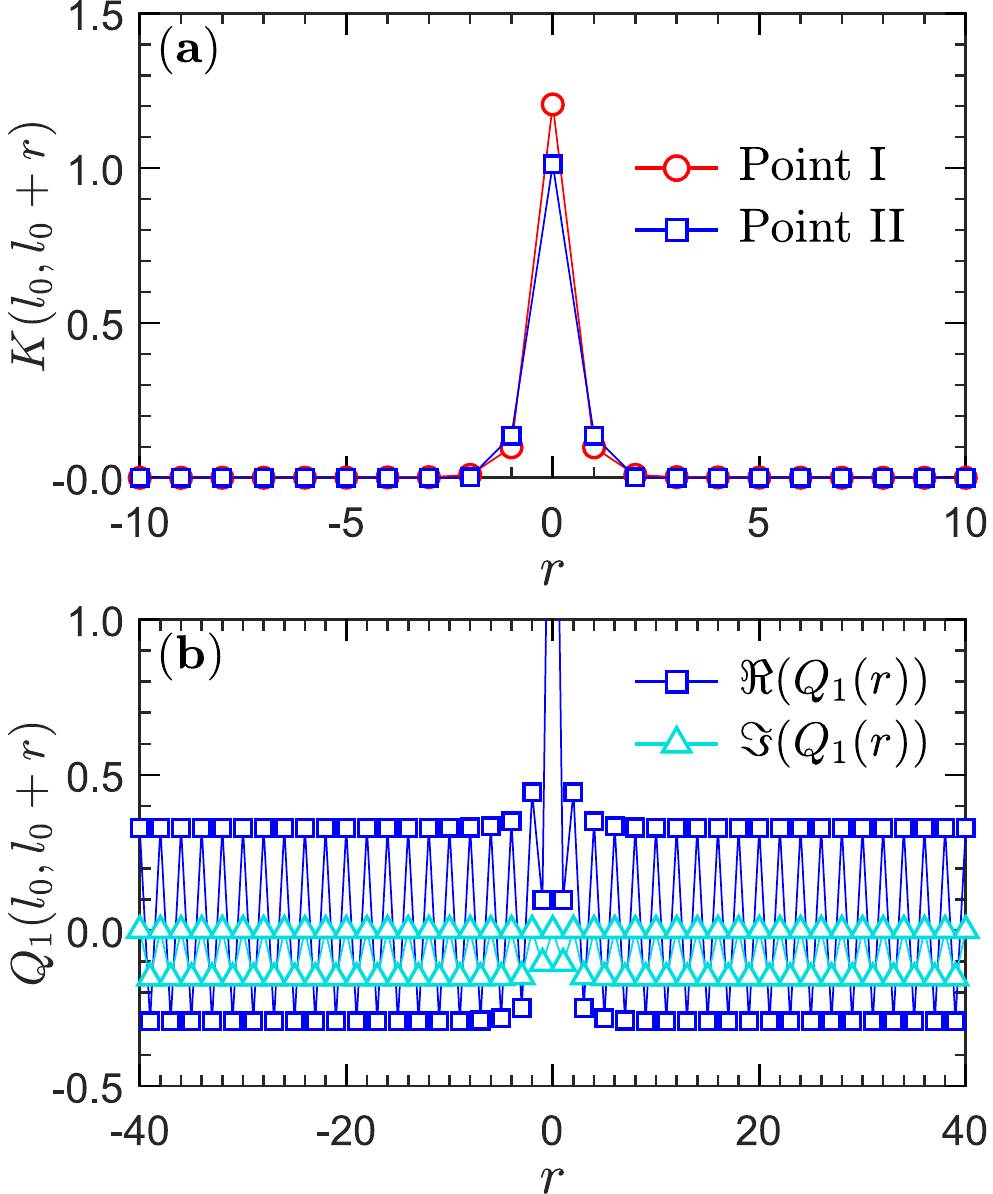}\\
  \caption{The (a) chiral-chiral correlation function $K(l_0, l_0+r)$ and (b) four-spin correlation function $Q_1(l_0, l_0+r)$ at selected data points in an open chain with the length of $L = 128$. Here, $l_0 = L/2$ and $r$ represents the distance from the reference point.
  In panel (a), the parameters are $D = -0.2$ and $\vartheta = \tan^{-1}(\sqrt2)$ for the point I and $D = -0.3$ and $\vartheta/\pi = 0.05$ for the point II.
  In panel (b), the parameters are equal to these of point I.
  }\label{FIG-VecChi}
\end{figure}

\begin{figure*}[htb]
\centering
\includegraphics[width=0.90\linewidth, clip]{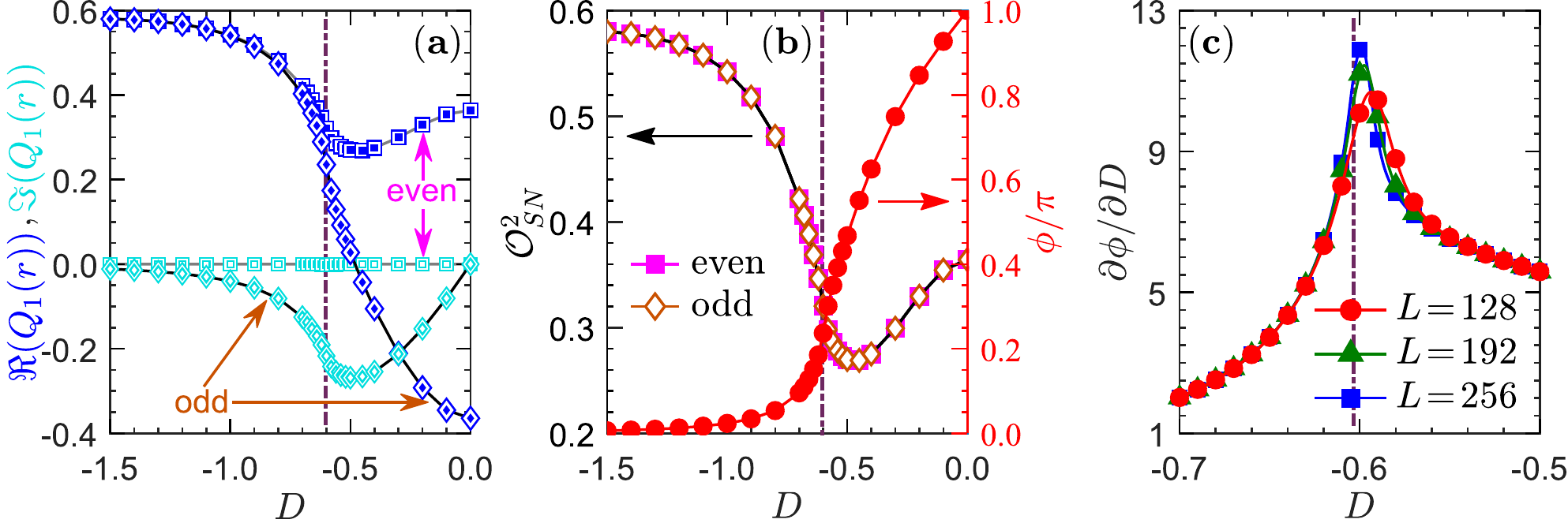}\\
\caption{Analysis of the spin-nematic correlation $Q_1(r)$ [defined in Eq.~\eqref{EQ:SNOP}] as a function of $D$ in the Kitaev spin chain with $\vartheta = \tan^{-1}(\sqrt2)$.
  In panels (a) and (b), the chain length is fixed at $L = 128$.
  (a) The real part $\Re(Q_1(r))$ (blue symbols) and imaginary part $\Im(Q_1(r))$(cyan symbols) of the spin-nematic correlation, with $r$ being even ($\square$) or odd ($\Diamond$).
  (b) The left axis shows the intensity $\mathcal{O}_{SN}^2$ of the spin-nematic correlation with $r$ being even (filled square) or odd (open diamond), while the right axis represents the nontrivial phase factor $\phi$ of the spin-nematic correlation when $r$ is odd.
  (c) The first-order derivative of the phase factor $\partial\phi/\partial D$ for a chain of length $L = 128$ (red circle), 192 (green triangle), and 256 (blue square).
  }\label{FIG-SpinNmtc}
\end{figure*}

\subsection{Spin-nematic phase}

The spin-nematic order is an intriguing phase which lacks the conventional magnetic order but breaks the spin-rotational symmetry,
giving rise to a nonzero quadrupolar order and possessing unusual excitations \cite{Andreev1984,Chandra1991,Chubukov1991}.
Therefore, emergence of the spin-nematic order is often related to the geometrical frustration and competing interactions which enhance quantum fluctuations \cite{Tsunetsugu2006,Kohamaa2019PNAS}.
Hitherto, several different scenarios have been proposed to theoretically realize the spin-nematic phase.
The spin-$1/2$ ferromagnetic chain with frustrated next-nearest-neighbor interaction is perhaps the most realistic model
since it is believed to characterize a couple of quasi-one-dimensional magnets like LiCuVO$_4$ \cite{Mourigal2012PRL,Buttgen2014PRB}.
According to the proposal by Zhitomirsky and Tsunetsugu \cite{Zhitomirsky2010EPL},
just below the saturation field, the gapped magnon excitations and the attractive interaction between them enforce the energy of the two-magnon bound state
is lower than that of the single-magnon state, thereby favoring the spin-nematic phase \cite{Hikihara2008PRB,Sudan2009PRB,Arlego2011PRB,Syromyatnikov2012,Parvej2017PRB}.
Theoretical analysis and numerical calculation suggest that the spin-nematic phase can be stabilized in spin-1 chains with the biquadratic interaction \cite{ManmanaPRB2011}.
In addition, the spin-nematic phase is also demonstrated to manifest itself in spin-1 chains whose Hamiltonians do not have $U(1)$ symmetry \cite{ZvyaginPRB2019}.

We start by checking for the possible existence of vector spin chirality
$\hat{\kappa}_i = (\textbf{S}_i \times \textbf{S}_{i+1})_z = -\imath (S_i^{+}S_{i+1}^{-} - S_i^{-}S_{i+1}^{+})/2$,
which is the vector product of two adjacent spins along the chain \cite{Chubukov1991,Hikihara2008PRB,Sudan2009PRB,Parvej2017PRB}.
The chiral order preserves the time-reversal symmetry but breaks the inversion symmetry.
The chiral-chiral correlation function is defined as
\begin{align}\label{EQ:VecChi}
K\big(i, j\big) = \langle\hat{\kappa}_i\hat{\kappa}_j\rangle,
\end{align}
in which $i$ and $j$ are site indices and we assume that $r \equiv \left|j-i\right|\rightarrow\infty$.
For concrete, we set $(i, j) = (l_0, l_0 + r)$ with $ l_0 = L/2$ and calculate the correlator $K\big(l_0, l_0+r\big)$ at two representative points, see Fig.~\ref{FIG-VecChi}(a).
It is observed that $K\big(l_0, l_0+r\big)$ decays rapidly with the distance $r$ and tends to zero, indicating that the chiral order is not favored in the ground state.
On the other hand, the spin-nematic order can be confirmed by the spin-nematic order parameter $\mathcal{O}_{SN}$,
which is extracted from the four-spin correlation function \cite{Syromyatnikov2012,Sato2013PRL}
\begin{align}\label{EQ:SNOP}
Q_{\delta}\big(i, j\big) = \langle S_{i}^{+} S_{i+\delta}^{+} S_{j}^{-} S_{j+\delta}^{-}\rangle \simeq \mathcal{O}_{SN}^2 e^{-\imath \phi}.
\end{align}
Here, $\delta$ is fixed as 1 throughout the paper, and $\phi$ is a phase factor that varies as the interaction strength changes.
The real (blue color) and imaginary (cyan color) parts of $Q_1(r) = Q_1\big(l_0, l_0+r\big)$ at a specific point $Q_1\big(l_0, l_0+r\big)$
in which $D = -0.2$ and $\vartheta = \tan^{-1}(\sqrt2)$ are shown in Fig.~\ref{FIG-VecChi}(b).
It is observed that depending on the odevity of $r$, $Q_1(r)$ has a strong even-odd effect.
When $r$ is even, $Q_1(r)$ is real as ${\Im}\big(Q_1(r)\big)$ is vanishingly small.
By contrast, both ${\Re}\big(Q_1(r)\big)$ and ${\Im}\big(Q_1(r)\big)$ saturate to finite values for odd $r$.
In any circumstance, the fact that the spin-nematic order parameter $\mathcal{O}_{SN}$ is nonzero manifests the existence of the spin-nematic order.
Of note is that the spin-rotational symmetry pertaining to the spin-nematic order is explicitly broken in the Hamiltonian.

We proceed to focus on the Kitaev chain with a [111]-type SIA to study the behavior of the spin-nematic order parameter.
The real (blue color) and imaginary (cyan color) parts of $Q_1(r \gg 1)$ for a chain of length $L = 128$ are shown in Fig.~\ref{FIG-SpinNmtc}(a).
Irrelevant of the strength of $|D|$, ${\Im}\big(Q_1(r)\big)$ vanishes when $r$ is even,
and is finite except for the limit case where $D = 0$ and an accidental point with $D \approx -0.49$ when $r$ is odd.
In the former case the phase factor $\phi$ is 0 while in the latter case it is nontrivial.
The left axis of Fig.~\ref{FIG-SpinNmtc}(b) illustrates the amplitude of $Q_1(r)$ when $r$ is even (pink square) and odd (brown diamond), respectively.
The fact that all the data points are overlapped indicates that $\mathcal{O}_{SN}^2$ is uniformly distributed and can be safely extracted from either case.
The right axis of Fig.~\ref{FIG-SpinNmtc}(b), on the other hand, shows the behavior of the phase factor $\phi$ as $D$ changes.
It decreases from $\pi$ in the pure Kitaev limit where $D$ is zero to $0$ when $|D|$ is large enough such that the system is in the deep AFM phase.
A nontrivial observation is that the phase factor $\phi$ undergoes a rapid change near the quantum critical region,
indicating that it may serve as a tool to probe the QPT.
To reveal the relation between phase factor $\phi$ and quantum criticality,
we show the derivative of $\phi$ with respect to tuning parameter $D$ in Fig.~\ref{FIG-SpinNmtc}(c).
The quantity $\partial\phi/\partial D$ displays a singular peak in the vicinity of the quantum critical point $D_c$,
with the height of peak growing and the position of peak approaching $D_c$ as the chain length $L$ increases.
Thus, $\partial\phi/\partial D$ is predicted to diverge as $L\to\infty$ and should in principle display a scaling behavior.
We note in passing that derivative of the geometric Berry phase associated with the many-body ground state
has already been demonstrated to exhibit universality in the neighbor of the quantum critical point \cite{ZhuPRL2006}.

\begin{figure}[!ht]
\centering
\includegraphics[width=0.95\columnwidth, clip]{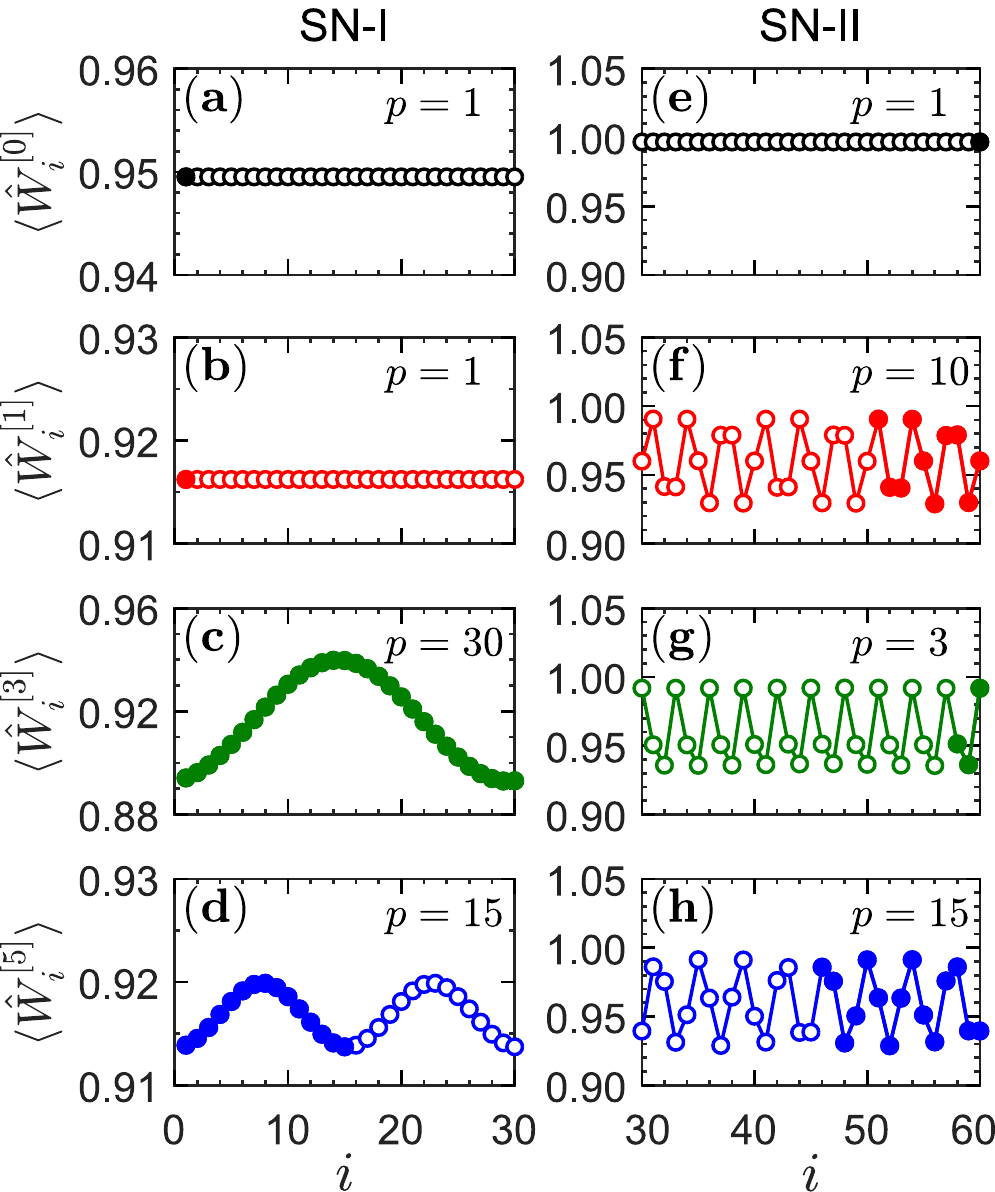}\\
\caption{The spatial pattern of the bond-parity operator $\langle W_i^{[l]}\rangle$ at site $i$ for the $l$-th energy level with $l = 0, 1, 2, \cdots$.
  The symbol $p$ marked in each panel stands for the periodicity of $\langle W_i^{[l]}\rangle$ in a chain of length $L = 60$.
  Panels (a)-(d) represent the spin-nematic phase of the type-I with $D = -0.3$ and $\vartheta/\pi = 0.30$,
  while Panels (e)-(g) represent the spin-nematic phase of the type-II with $D = -0.3$ and $\vartheta/\pi = 0.05$.
  }\label{FIG-SNWp}
\end{figure}

Whereas the spin-nematic phase is characterized by a unique ground state under PBC, its excited states are quite involved and display distinct patterns.
We find that all the excited states are doubly degenerate except for the first excited ground state.
The first excited ground state is unique in the wide region, as compared to the twofold case observed in a specific area where $|D|$ and $\vartheta$ are small.
Therefore, we distinguish the spin-nematic phase as type-I and type-II, respectively,
based on its degeneracy of the first excited state (for illustration, see Fig.~\ref{FIG-GSPD}).
However, since the lowest excitation gap of the spin-nematic phase does not close throughout its whole region,
there is not a QPT but a likely crossover between the two.
To illustrate it, we have calculated the phase factor $\phi$ at fixed SIA, saying $D = -0.3$.
The derivative of $\phi$ with respect to $\vartheta$ shows a broad hump and suffers from an insignificant finite-size effect,
characteristic of crossover phenomenon.

To further discriminate the two different types of spin-nematic phase,
we resort to the bond-parity operator $\hat{W}_i$ defined as \cite{SenShankar2010,YouSunRen2020}
\begin{equation}\label{EQ:WbEvOd}
\hat{W}_{2i-1} = \Sigma_{2i-1}^{y}\Sigma_{2i}^{y},\quad
\hat{W}_{2i}   = \Sigma_{2i}^{x}\Sigma_{2i+1}^{x},
\end{equation}
where $\Sigma_i^{\alpha} = e^{\imath\pi S_i^{\alpha}}$ is the on-site operator.
For the pure Kitaev chain, $\hat{W}_i$ commutes with the Hamiltonian such that its eigenvalues should only be $\pm1$ for the ground state.
By switching on the SIA, the relation $\big[\hat{W}_i, \mathcal{H}\big] = 0$ does not hold as long as $\vartheta \neq 0$,
indicating that $\langle\hat{W}_i\rangle$ will deviate from 1.

The spatial patterns of $\langle W_i^{[l]}\rangle$ in a closed chain of $L = 60$ at different energy levels $l = 0, 1, 3, 5$
for the type-I and type-II spin-nematic phases are shown in Fig.~\ref{FIG-SNWp},
with $D = -0.3$ and $\vartheta/\pi$ = $0.30$ and $0.05$ for the left and right panels, respectively.
It can be seen from Fig.~\ref{FIG-SNWp}(a) and Fig.~\ref{FIG-SNWp}(e) that the ground-state patterns of $\langle W_i^{[0]}\rangle$ for both types
are uniformly distributed with a periodicity $p = 1$ along the chain.
For the excited-state patterns, they display a similarity within the twofold degenerate states and thus only three selected energy levels are shown.
For the spin-nematic phase of the type-I, the first excited state is again unique and $\langle W_i^{[1]}\rangle$ is completely flat.
$\langle W_i^{[3]}\rangle$ and $\langle W_i^{[5]}\rangle$ are smoothly changed within the chain,
with periodicity $p = 30$ (see Fig.~\ref{FIG-SNWp}(c)) and $p = 15$ (see Fig.~\ref{FIG-SNWp}(d)), respectively.
By contrast, while $\langle W_i^{[l]}\rangle$ ($l = 1, 3, 5$) exhibits periodicity of $p = 10$ (see Fig.~\ref{FIG-SNWp}(f)), $p = 3$ (see Fig.~\ref{FIG-SNWp}(g)), or $p = 15$ (see Fig.~\ref{FIG-SNWp}(h)), its values are quite fluctuating and elusive.
However, pertaining to the behavior of the first excited state,
the flatness versus oscillation of $\langle W_i^{[1]}\rangle$ is the hallmark of the difference between the type-I and type-II spin-nematic phases.
It is in this sense that we can identify the crossover boundary of the two by the standard deviation of $\langle W_i^{[1]}\rangle$, i.e., $\sigma_W$.
In our calculation on three closed chains of length $L$ = 24, 48, and 72,
the quantity $\sigma_W$ undergoes a sharp jump at $\vartheta/\pi \approx 0.13$, as depicted in Fig.~\ref{FIG-GSPD}.
We note that the periodicity of $\langle W_i^{[l]}\rangle$ in the excited states should be different as we change the chain length,
and such a periodicity can be discerned by the discrete Fourier transform of $\langle W_i^{[l]}\rangle$.
Nevertheless, the most remarkable feature that the curves of $\langle W_i^{[l]}\rangle$ ($l > 0$) are smooth and discrete, respectively,
in the type-I and type-II spin-nematic phases remains preserved.

Finally, we comment on the mechanism of the spin-nematic phase.
Hitherto, the two-magnon bound state picture in frustrated spin-$1/2$ systems with the nearly saturated magnetic field
and the description of the on-site quadrupolar order in spin-1 models with the biquadratic interaction
are widespread to describe the spin-nematic phase.
More interestingly, an attempt to unify these scenarios based on the language of spin-1 dimers has been proposed \cite{Tanaka2020PRB}.
Physically, the presence of magnetic step of two in magnetization curve or the Anderson tower of states containing only the even total spin sectors \cite{Lauchli2005PRL}
is known as the fingerprint of the spin-nematic phase.
However, it seems to be infeasible to check the picture as the total spin is not a conserved quantity for the lack of $U(1)$ symmetry.
Nevertheless, one can calculate the one-magnon and two-magnon dynamical spectra, from which the magnon and magon-pair gaps can be extracted.
This may give some clues on the nature of the spin-nematic phase and deserves future study.

\begin{figure*}[htb]
\centering
\includegraphics[width=0.90\linewidth, clip]{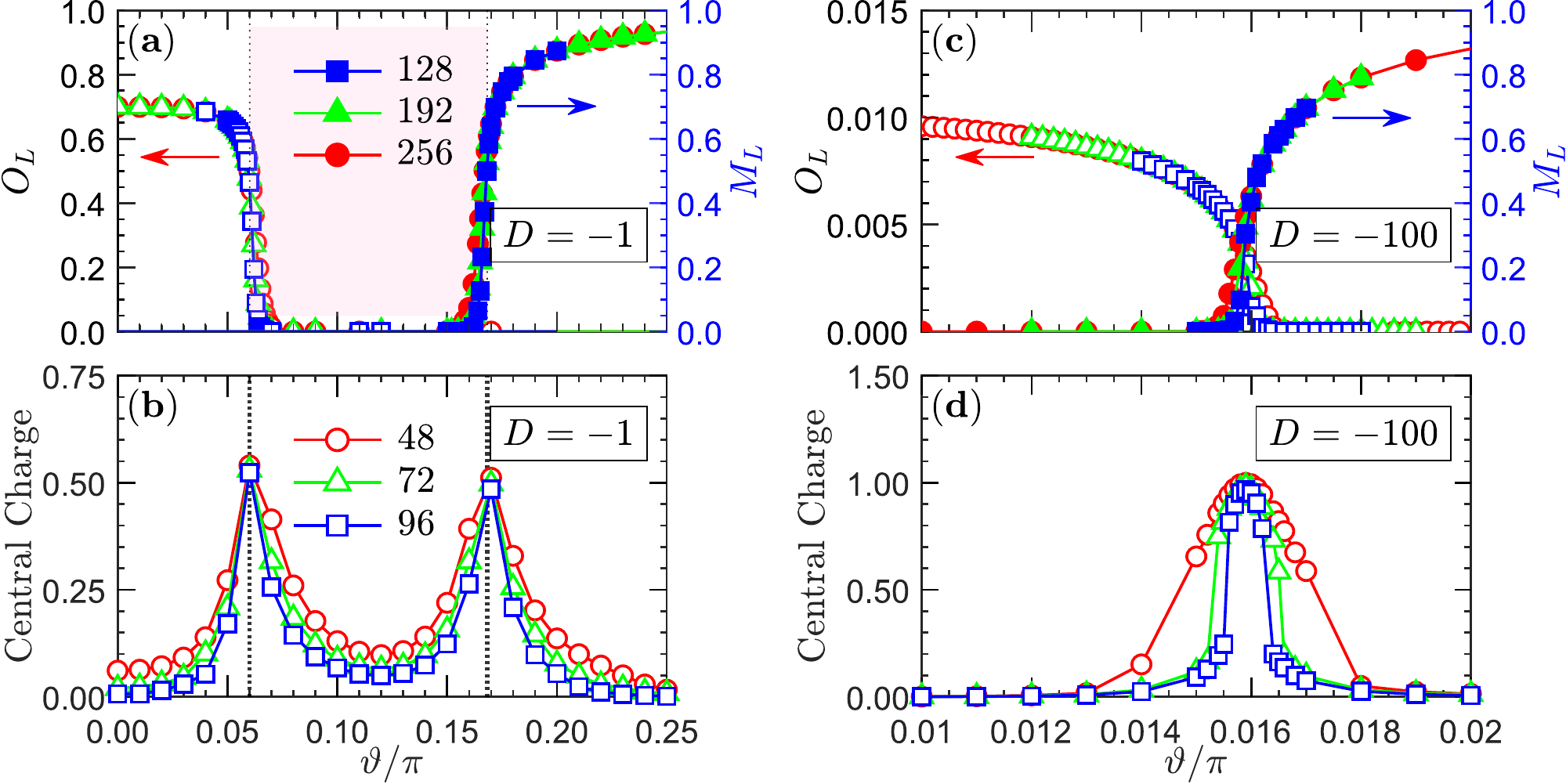}\\
\caption{(a) and (b) show behaviors of order parameters and central charge as a function of $\vartheta/\pi$ in the Kitaev spin chain with $D = -1$.
  (a) The dimerized order parameter $O_L$ (left axis, open symbols) and AFM order parameter $M_L$ (right axis, filled symbols) for an open chain of length $L = 128$ (red circle), 192 (green triangle), and 256 (blue square). The intervening pink region represents the spin-nematic phase.
  (b) The central charge for a periodic chain of length $L = 48$ (red circle), 72 (green triangle), and 96 (blue square).
  (c) and (d) are the same as these of (a) and (b) but for $D = -100$, in which a direct QPT between the dimerized phase and AFM phase occurs.
  }\label{FIG-DQCP}
\end{figure*}

\subsection{Deconfined quantum critical point}

Dating back to 2004, the DQCP is a fascinating proposal which asserts a continuous QPT between two spontaneous symmetry-breaking phases with completely unrelated broken symmetries \cite{Senthil2004Science}.
Right at the DQCP,  deconfined fractionalized particles appear, accompanying by an emergent symmetry to reconcile the two different order parameters nearby.
This scenario is clearly beyond the conventional LGW paradigm as the latter predicts that this kind of QPT should be of first order.
While the transition between the AFM phase and the valence-bond-solid phase in two dimension is regarded as the possible realization of the deconfined criticality,
decisive evidences are still lacking as a weakly first-order QPT cannot be ruled out \cite{Sandvik2007}.
The one-dimensional analogy was put forward in 2019, providing another feasible way towards unraveling the enigmatic DQCP \cite{JiangMotrunich2019}.
Massive numerical work has been devoted to studying the DQCP in one-dimensional spin-$1/2$ models during the past few years,
including the ferromagnetic frustrated spin chain \cite{Roberts2019PRB,Huang2019PRB,Sun2019PRB,Luo2019PRB},
the spin ladder with ring-exchange interaction \cite{Ogino2021PRB},
and the Kitaev spin chain with multiple-spin interaction \cite{Macedo2022}.

We will demonstrate that the spin-1 Kitaev chain with tunable SIA is another promising platform that exhibits the DQCP.
To begin with, we focus on the line of $D = -1$ and calculate the dimer order parameter $O_L$ and magnetic order parameter $M_L$, see Fig.~\ref{FIG-DQCP}(a).
It can be seen that both order parameters decrease smoothly as the driving parameter $\vartheta$ approaches their corresponding quantum critical points.
In the intervening region where $0.0601 \lesssim \vartheta/\pi \lesssim 0.1683$, the two order parameters vanish and the spin-nematic phase of type-I survives,
in accordance with the fact that the spin-nematic phase preserves the translational symmetry and time-reversal symmetry.
Next, we appeal to the central charge to pin down the nature of QPTs.
The central charge $c$ is usually extracted from the entanglement entropy which is known to obey the conformal field theory \cite{CalCar2004}.
Although the OBC is frequently adopted in the DMRG calculation,
it can induce an intrinsic alternating term which decays away from the boundary with an approximately power-law behavior in the entanglement entropy,
making the fitting formula more intricate \cite{Laflorencie2006PRL}.
Therefore, we turn to the PBC and the entanglement entropy is well described by the following expression \cite{CalCar2004}
\begin{equation}\label{EQ:VNECC}
\mathcal{S}_{L}(x) = \frac{c}{3}\ln\left[\frac{L}{\pi}\sin\Big(\frac{\pi x}{L}\Big)\right] + c',
\end{equation}
where $x$ is the length of a subsystem and $c'$ is a nonuniversal constant.
Results of the fitted central charge for three different lengths $L$ are shown in Fig.~\ref{FIG-DQCP}(b).
It is found that at $\vartheta/\pi \approx 0.0601$ and $\vartheta/\pi \approx 0.1683$ the central charges are slightly decreases with the increase of the system
but saturate to $1/2$ eventually, indicating that both QPTs belong to the Ising universality class.

As the intensity of the SIA increases, region of the spin-nematic phase shrinks slightly and does not disappear until $|D|$ is large enough.
After a careful inspection of the quantum criticality, we take $D = -100$ as an example to illustrate the direct QPT between the dimerized phase and the AFM phase.
The behaviors of order parameters $O_L$ and $M_L$ in a narrow window of $0.01 \leq \vartheta/\pi \leq 0.02$ are shown in Fig.~\ref{FIG-DQCP}(c).
They are smoothly changed as $\vartheta$ varies and the finite-size scaling [see Eq.~\eqref{EQ:FSS}] suggests that there is only a sole quantum critical point at $\vartheta/\pi \approx 0.0158$.
In Fig.~\ref{FIG-DQCP}(d), we also fit the central charge in the same parameter range as that of Fig.~\ref{FIG-DQCP}(c).
Far away from the critical region, the central charge is vanishingly small and tends to be zero with the increase of the system size, indicative of the gapped ground states.
In the critical region, the central charge is sizable and its maximal value is extremely close to 1.
Such a finite central charge is also confirmed in several independent calculations like $D = -200$.
Since the nonzero central charge is crucial to corroborate the continuous QPT,
our result thus demonstrates that the dimer-AFM transition is continuous.
Nevertheless, determining the nature of this QPT is numerically challenging,
albeit a conceivable possibility is the Gaussian transition which has been proposed in other similar situations \cite{Mudry2019PRB}.
Notably, because the broken translational symmetry and dihedral symmetry are totally irrelevant,
the continuous QPT is forbidden by the LGW paradigm and thus the quantum critical point is interpreted as a DQCP.

\section{Conclusion}\label{SEC:CONC}

We have studied the quantum phase diagram of a spin-1 Kitaev chain with tunable SIA by the DMRG method,
which is identified to host a dimerized phase, an AFM phase, and two distinct spin-nematic phases.
In line with the previous research effort which reveals that the ground state of the spin-1 Kitaev chain is a nonmagnetic Kitaev phase \cite{LuoPRR2021},
we further clarify that it is a spin-nematic order which preserves the translational symmetry and time-reversal symmetry but breaks spin-rotational symmetry,
giving rise to a finite spin-nematic correlation.
The four-spin correlation function pertaining to the spin-nematic order parameter can exhibit a nontrivial phase factor that varies as the SIA $|D|$ changes,
and the derivative of the phase factor is demonstrated to be a useful probe to capture QPTs.
Depending on the degeneracy of the first excited state,
the spin-nematic phase can be classified into two types and the model undergoes a crossover between the two.
Notably, the nature of the spin-nematic phase is an intriguing topic which deserves future study.

As the strength of SIA increases, the dimerized phase and the AFM phase with broken translational symmetry and dihedral and time-reversal symmetries set in
when the SIA is aligned along the [001] direction and [111] direction, respectively.
Of particular note is that the spontaneous dimerization is induced by the SIA only,
highlighting the unique role played by the Kitaev interaction.
When the SIA is modest, the spin-nematic phase is intervened between the two spontaneous symmetry-breaking phases,
and both QPTs belong to the Ising universality class.
By contrast, the spin-nematic phase is destroyed by strong SIA, leading to a continuous QPT between the dimerized phase and the AFM phase.
Thus, our result demonstrates that the Kitaev-type spin chain can offer a promising playground to study the DQCP.
In the future, it is desirable to study the emergent symmetry \cite{Huang2019PRB}, the dynamic signatures \cite{Xi2020CPL},
the fidelity and entanglement from the quantum information aspect \cite{Sun2019PRB,Yang2021PRE},
and the nonequilibrium critical dynamics described by Kibble-Zurek mechanism \cite{Huang2020PRR}
in the critical region so as to corroborate this exotic QPT.

\begin{acknowledgments}
This work is supported by the National Program on Key Research Project (Grant No. MOST2022YFA1402700),
the Natural Science Foundation of Jiangsu Province (Grant No. BK20220876),
the National Natural Science Foundation of China (Grants No. 12247183, No. 12274187, No. 12247101, No. 12174020, No. 11974244, and No. U1930402),
and the NSERC Discovery (Grant No. 2022-04601).
Q.L. also acknowledges the Fundamental Research Funds for the Central Universities (Grant No. NS2022097)
and the startup Fund of Nanjing University of Aeronautics and Astronautics (Grant No. YAH21129).
H.-Y.K. also acknowledges funding from the Canadian Institute for Advanced Research and the Canada Research Chairs Program.
The computations are partially supported by High Performance Computing Platform of Nanjing University of Aeronautics and Astronautics (NUAA)
and Tianhe-2JK at the Beijing Computational Science Research Center (CSRC).
Computations are also performed on the Niagara supercomputer at the SciNet HPC Consortium.
SciNet is funded by the Canada Foundation for Innovation under the auspices of Compute Canada,
the Government of Ontario, Ontario Research Fund, Research Excellence, and the University of Toronto.
\end{acknowledgments}




%

\end{CJK*}

\end{document}